\documentclass[
 ,final            
  ]
  {aipproc}

\layoutstyle{6x9}

\begin{document}

\title{Summary of the Structure Functions and Low-x working group}


\author{Joanne E. Cole}{
  address={H.H. Wills Physics Laboratory,
  University of Bristol, Bristol, BS8 1TL, UK}
}

\author{Jianwei Qiu}{
  address={Department of Physics and Astronomy
Iowa State University, Iowa 50011, USA}
}

\author{Un-ki Yang}{
  address={Enrico Fermi Institute, University of Chicago,Chicago,
  Illinois 60637, USA}
}

\begin{abstract}
We report a summary of the structure function working group which
covers a wide range of the recent results from HERA, Tevatron,
RHIC, and JLab experiments, 
and many theoretical issues from low $x$ to high $x$.
\end{abstract}

\maketitle


\section{Introduction}
Much of the predictive power of Quantum Chromodynamics (QCD) is
provided by universality of the non-perturbative functions, in
particular, the parton distribution functions (PDFs), in
factorization theorems for hard processes.  With the aid of
factorization and perturbative calculation of short-distance dynamics,
the universality allows us to extract a set of PDFs from some
reactions and then use them to predict observables in other reactions.     
Knowledge of PDFs is critical for testing QCD dynamics in
asymptotic region at existing facilities, as well as, for making
predictions for future facilities, like the Large Hadron Collider
(LHC).  It is also essential for exploring non-perturbative QCD
dynamics when the extracted PDFs are compared with what calculated in
lattice QCD or in effective field theory approaches. 

With only one identified hadron, structure functions of inclusive
lepton-hadron deeply inelastic scattering (DIS) are clean observables
for extracting PDFs.  After more than 30 years of continuous effort,
and many generations of machines and detectors, measurements of proton
structure functions have become the benchmark tests of QCD dynamics.
With the HERA at DESY, we are able to explore the kinematic region
with the Bjorken $x_B$ as low as $10^{-5}$ while staying the DIS
regime.  The continuous growth of structure functions as $x_B$
decreases raises an urgent question: when such growth will hit the
unitarity limit and slow down?  
The knowledge of structure functions and
low $x$ physics is extremely important for testing QCD dynamics and
our ability to explore new physics beyond the standard model.

Our working group had a total of 46 talks divided into 9 sessions
including one joint session with Electroweak and Beyond the Standard
Model working group.  
In this writeup, we summarize the recent achievements, progresses, and
open questions that were presented at our working
group meetings.  We organize this summary into seven parts:
\begin{enumerate}
\item Structure function measurements at low $x$
\item Structure functions and PDFs at high $x$
\item Progress in the determination of PDFs
\item Toward QCD precision tests
\item Low-$x$ physics: parton evolution and saturation
\item Nuclear structure functions and nuclear PDFs
\item New approaches to PDFs
\end{enumerate}

\section{Structure Function measurements at low $x$}

Measurements of the proton structure function, $F_2$, in neutral current
(NC) deep inelastic scattering (DIS) at HERA are vital for testing the
predictions of perturbative QCD and in the determination of the parton
distribution functions of the proton.  Recent results from the two 
general-purpose detectors, H1 and ZEUS, cover five orders of magnitude in the
photon virtually, $Q^2$, and in the Bjorken scaling variable, 
$x$~\cite{heraf2}.

The possibility afforded by HERA of studying the structure functions down to
values of $x$ as low as $\sim 10^{-6}$ is important, as it gives access to
partons which have undergone a large number of QCD branching processes.  The
density of these partons, both gluons and the so-called ``sea'' quarks has
be found to increase dramatically as $x$ decreases, which may indicate the
need to taken into account non-linear effects in QCD evolution, such as
saturation.

The double-differential cross section for inclusive NC DIS is given by:
\begin{equation}
\frac{xQ^4}{2 \pi \alpha^2 Y_+} \frac{d^2 \sigma}{dxdQ^2} = \sigma_r = F_2 (x,Q^2) - \frac{y^2}{Y_+} F_L (x,Q^2) - \frac{Y_-}{Y_+} xF_3 (x,Q^2)
\label{eq:dd}
\end{equation}
where $Y_{\pm} = 1 \pm (1 - y)^2$, in which $y = Q^2/xs$ is the inelasticity,
$s$ is the total squared center-of-mass energy and $F_2$, $F_L$ and $F_3$ are
the structure functions of the proton.  The quantity, $\sigma_r$ is also
defined in this equation, which is known as the reduced cross section.

Although the precision measurements of $F_2$ exist over such a large
kinematic range, the same cannot be said for the longitudinal structure
function, $F_L$, which has not been directly measured at HERA.  $F_L$ is
directly sensitive to scaling violations and hence to the gluon content of
the proton and is therefore crucial to our understanding of proton structure.

The final term in equation~\eqref{eq:dd} contains $xF_3$, or the 
parity violating structure function.  This structure function is, however,
only important at high $Q^2$ and will not be considered further in this
section, where the interest is primarily in low $Q^2$ structure function
measurements.

\begin{figure}
  \includegraphics[height=.5\textheight]{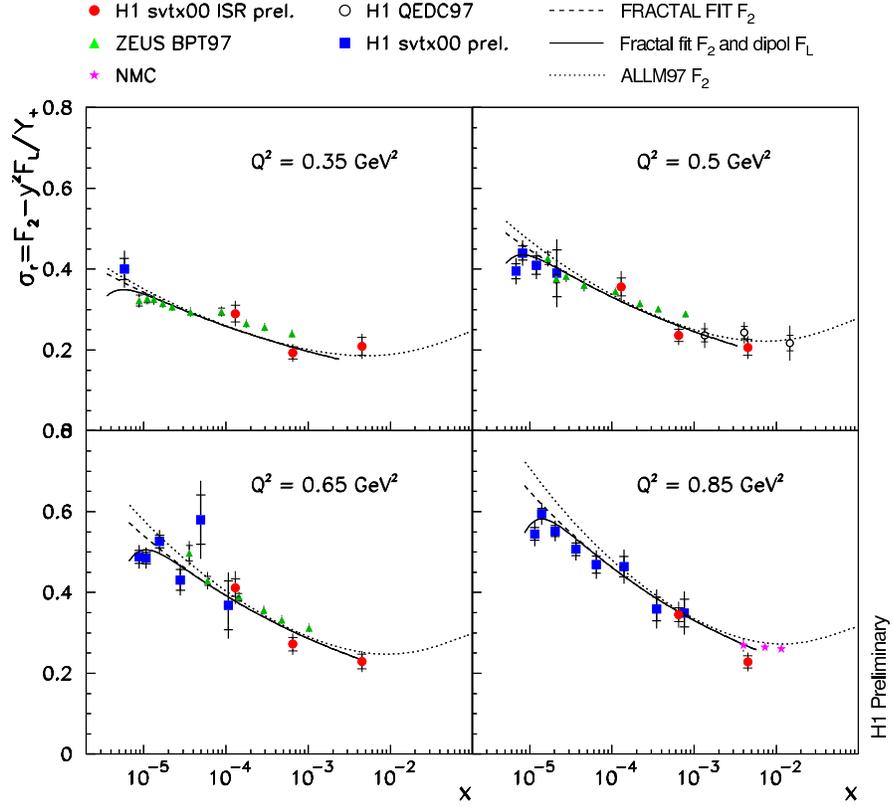}
  \caption{Compilation of reduced cross section measurements for $Q^2 < 1$ GeV$^2$ from the H1, ZEUS and NMC Collaborations.}
  \label{fig:lowq2}
\end{figure}
The rise of $F_2$ with decreasing $x$ persists down to very low values of
$Q^2$~\cite{bptf2}, although it is known that as $Q^2 \rightarrow 0$, 
$F_2 \rightarrow$ constant $Q^2$, as it must in order to satisfy the 
conservation of the electromagnetic current.  It is also known that around 
$Q^2 \sim 1$ GeV$^{2}$, perturbative QCD begins to breakdown and 
phenomenological models must be invoked to explain the behavior of the
$F_2$ data.

From an experimental point of view, accessing very low values of $Q^2$ is
technically challenging, but has been achieved by the HERA experiments using a 
number of different techniques.  The H1 Collaboration presented recent 
measurements of the reduced cross section at low $Q^2$ using two of these 
techniques, namely, via the identification of QED Compton events~\cite{qedc} 
and using a small sample of data in which the interaction vertex was 
intentionally shifted by $+70$ cm toward the outgoing proton beam 
direction~\cite{svtx}, effectively extending the acceptance of the H1 detector 
to values of $Q^2$ down as low as $0.35$ GeV$^{2}$.  Using this so-called 
``shifted vertex'' data sample, they have also specifically identified events 
in which an energetic photon was emitted by the incoming lepton prior to its 
interaction with the proton; these initial-state radiative (ISR) events give 
access not only to even lower values of $Q^2$, but also to higher values of 
$x$, giving a wide coverage in $x$ at low $Q^2$.

\begin{figure}
  \includegraphics[height=.4\textheight]{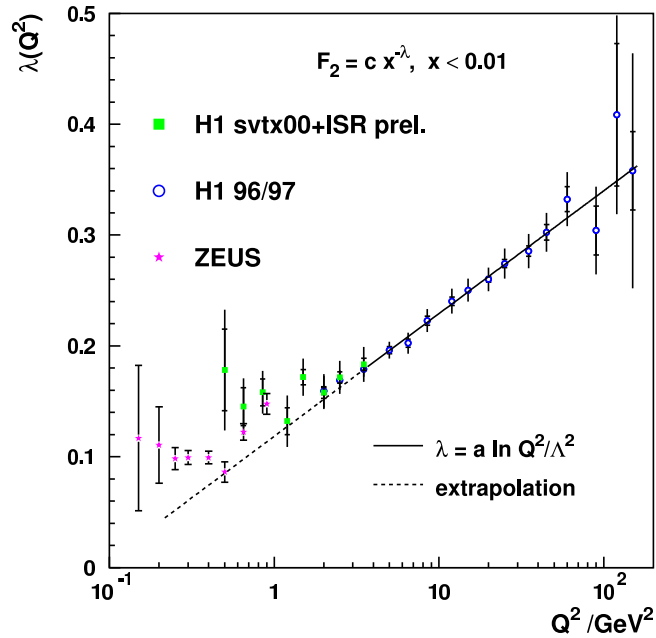}
  \caption{Compilation of selected HERA results on the parameter $\lambda$, obtained from fits of the form $F_2 = c(Q^2) \cdot x^{- \lambda(Q^2)}$ to low $x$ data}
  \label{fig:lam}
\end{figure}
These measurements are shown in figure~\ref{fig:lowq2}, in which it can be seen
that $F_2$, and hence the reduced cross section, rises with decreasing $x$, 
even at low values of $Q^2$.  The only exception to this behavior is at the 
very lowest values of $x$, at which the contribution to the reduced cross 
section from the longitudinal structure function, $F_L$, becomes significant,
causing $\sigma_r$ to decrease.  As can be seen in equation~\eqref{eq:dd}, the 
contribution to the reduced cross section from $F_L$ is suppressed for all but 
the highest values of $y$ (low $x$).  This behavior can be exploited to
perform an extraction of $F_L$, albeit in a model-dependent way.  The resulting
$F_L$ points were also shown at this workshop~\cite{svtx} and are already able 
to discriminate between different PDF parameterizations.

The low $Q^2$ $F_2$ data can be fitted in order to quantify the change of
the low $x$ slope of $F_2$ with $Q^2$.  Figure~\ref{fig:lam} shows the result 
of just such a fit, performed for $x < 0.01$ by the H1 Collaboration.  The 
expected change in behavior around $Q^2 \sim 1$ GeV$^2$, is clearly observed.

\section{Structure Functions and PDFs at High $x$}

Structure Functions at high $x$ region has brought many attentions
at this workshop. Recently, it has been realized that it is very important
to understand this region in order to achieve precise electroweak measurements
and to extract new physics signals from HERA, Tevatron and LHC at high $Q^2$ region.
A large uncertainty on the PDFs at very high $x$ and low $Q^2$ region 
can make a big impact on the high $Q^2$ region even at intermediate $x$ due to the effect of DGLAP evolution.

\begin{figure}
  \includegraphics[height=.4\textheight]{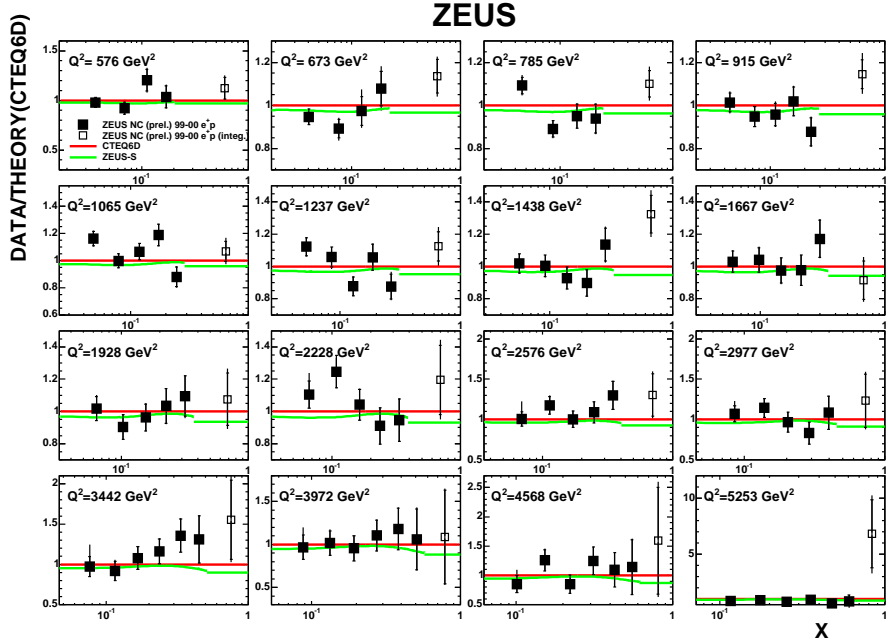}
  \caption{[Left] The ratio of the ZEUS differential cross sections data and predictions
  using CTEQ6D PDFs.}
  \label{fig:zeus_highx}
\end{figure}

Most precise data on high $x$ come from the traditional fixed target experiments
(SLAC/BCDMS/NMC). But their high $x$ data corresponds to low $Q^2$ region
where we face many challenges
in understanding all non-perturbative QCD and nuclear effects.
One clean way is to probe the structure functions at high $x$ and $Q^2$ directly.
Both H1 and ZEUS~\cite{heraQ2} showed measurements of the cross sections for neutral and charged-current 
scattering as a function of $Q^2$ using polarized beams. The measured cross sections
are well described by the Standard Model. But more data is required to constrain parton distributions functions at high $x$.
The ZEUS~\cite{ning}  presented a very promising method to probe the PDFs up to $x=1$
using the jet information ($E_{jet}$ and $\theta_{jet}$)
to calculate the value of $x$. Events with no jets reconstructed within their fiducial
volume is assumed to come from very high $x_{edge}$ to 1. The measured cross sections
using early dataset show good agreements with the predictions using CTEQ6D PDFs, shown 
in Figure~\ref{fig:zeus_highx}.
However, their highest $x$ data
tend to be higher than the predictions. Thus, it would be interesting to see their results
using a full dataset.

\begin{figure}
  \includegraphics[width=.5\columnwidth,height=.5\textheight]{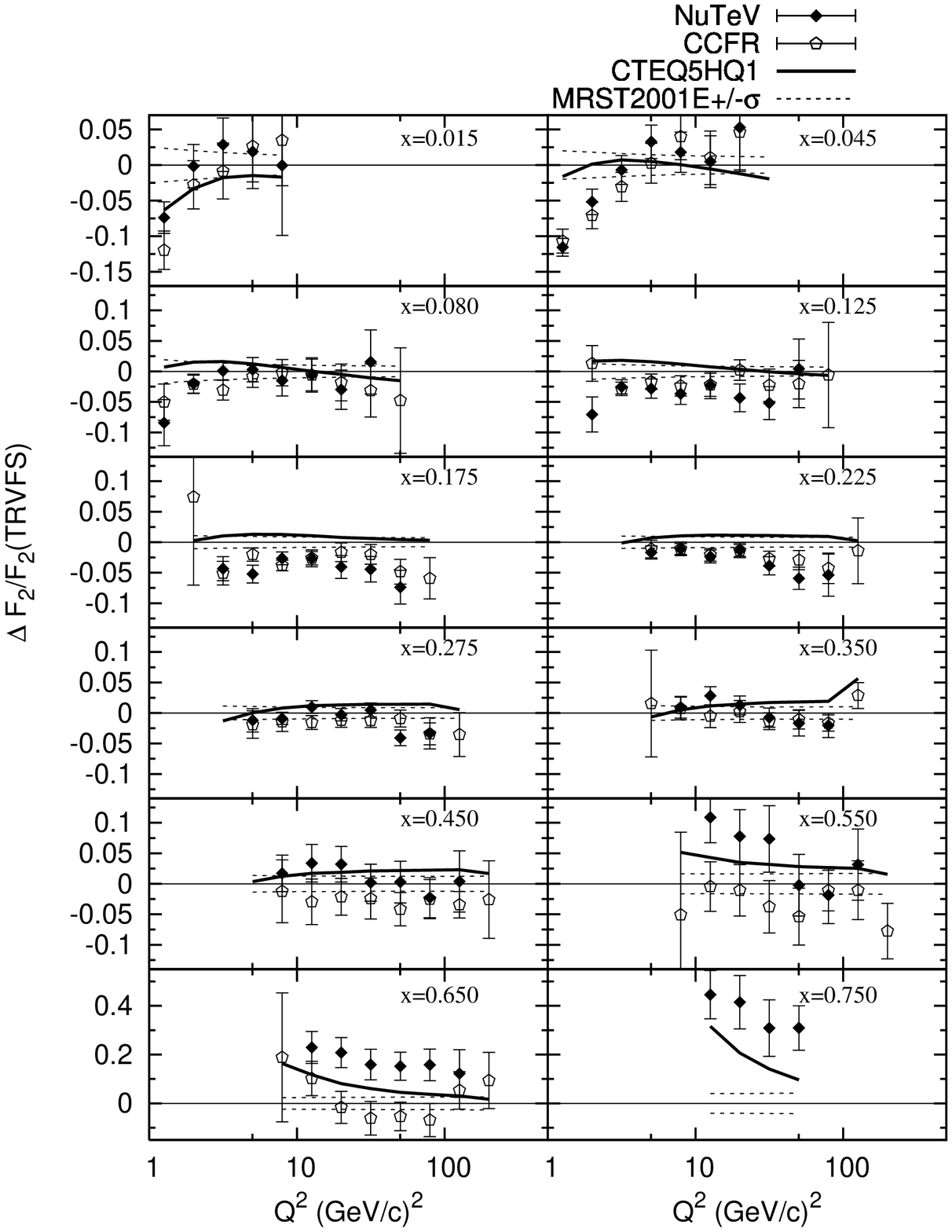}
  \includegraphics[width=.5\columnwidth,height=.5\textheight]{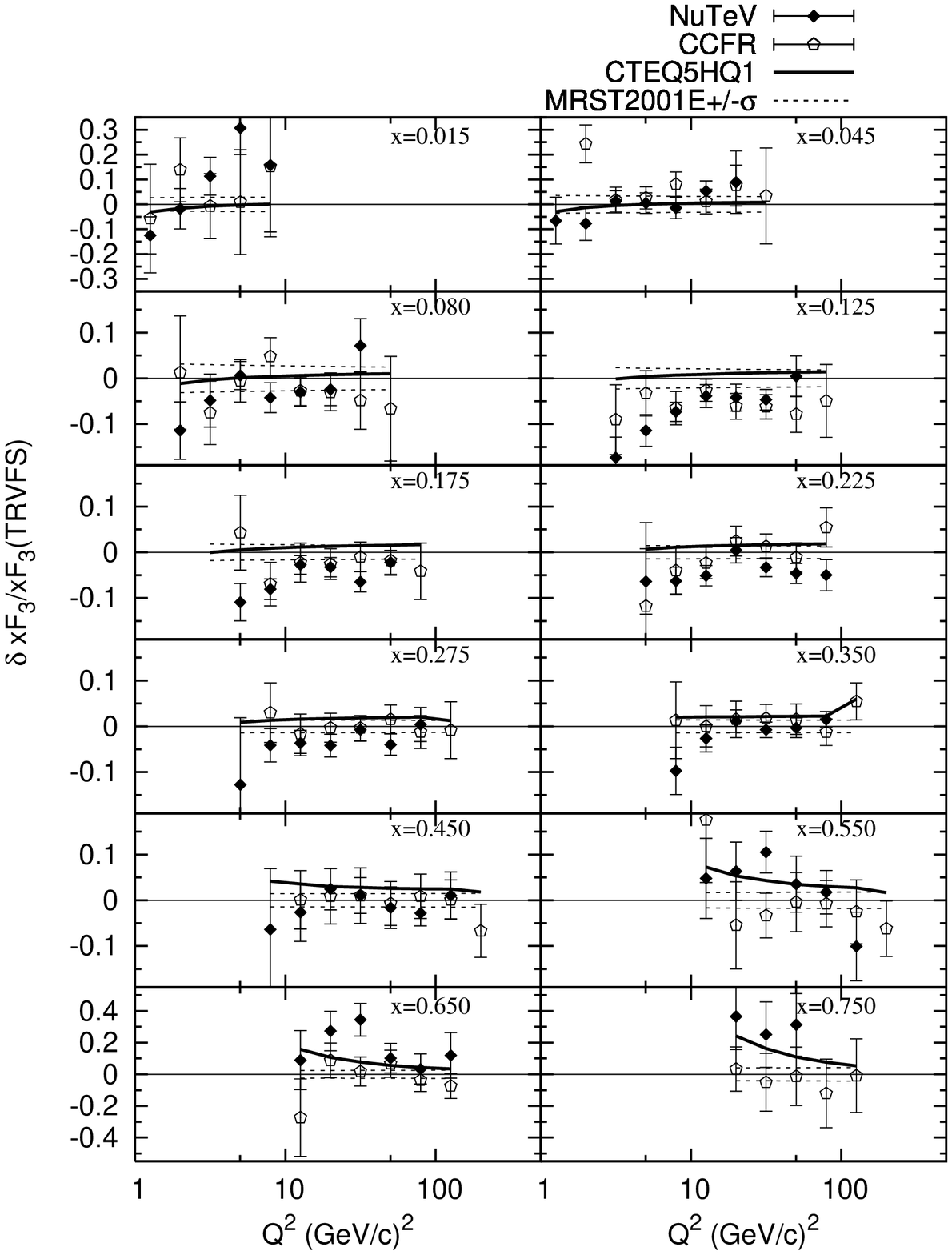}
  \caption{NuTeV $F_2$[left] and $xF_3$[right] data are compared with CCFR data 
  and the NLO prediction with MRST2001E.}
  \label{fig:f2_nutev}
\end{figure}

The NuTeV~\cite{martin} presented their final differential cross sections
using neutrino-iron scattering. The extracted $F_2$ and $xF_3$ from their differential cross sections
are 20\% higher than the CCFR measurements, and 10-15\% higher than the BCDMS $F_2$, as shown in Figure~\ref{fig:f2_nutev}.
They explained that two third of the difference between the NuTeV and CCFR measurements
is due to an improved calibration of the magnetic field, and a better modelling in Monte Carlo.
This result implies that that nuclear effect in neutrino scattering is different
from that in the charged lepton scattering at high $x$. Thus, we need to resolve this difference
before the NuTeV data can be used in a global PDFs analysis to constrain the PDFs at high $x$. It would be interesting to see their QCD fit results. A possible difference in the nuclear effect can be resolved by the CHORUS data on the lead target, and future Minerva/MINOS results.

\begin{figure}
  \includegraphics[width=.5\columnwidth,height=.37\textheight]{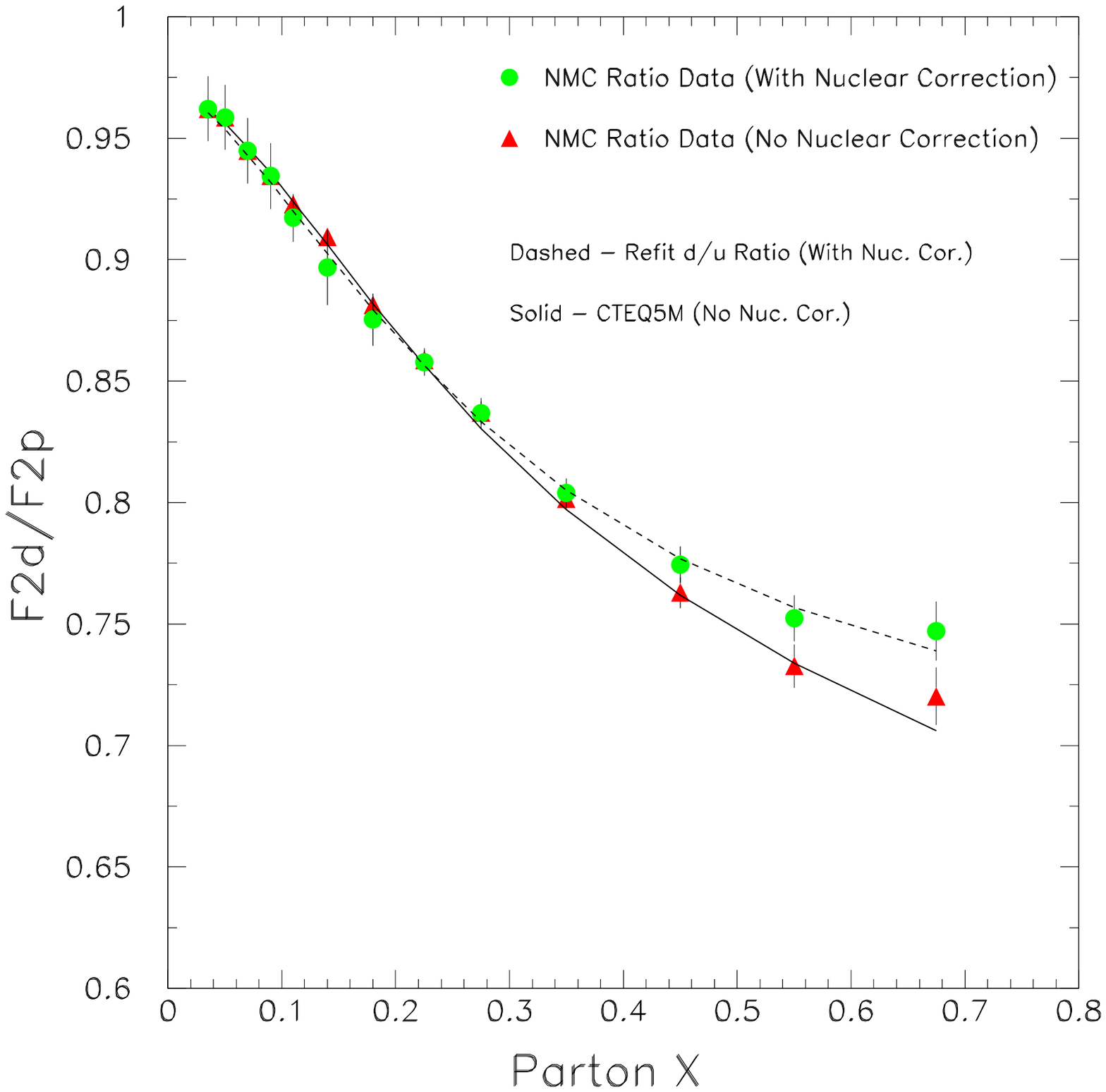}
  \includegraphics[width=.5\columnwidth,height=.35\textheight]{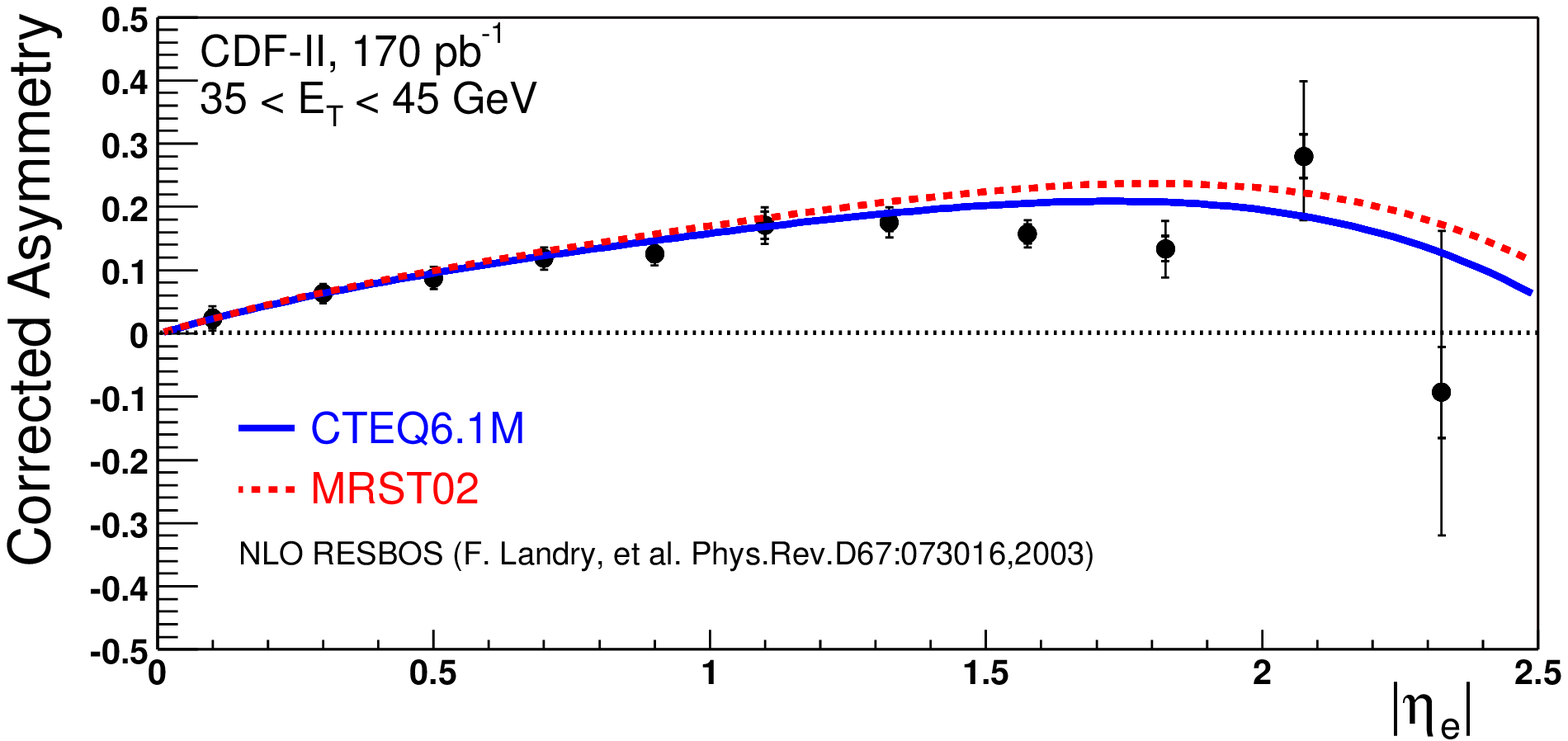}
  \caption{[Left] The ratio of $F_2$ data on the deuterium and hydrogen targets with and without
   the nuclear corrections. [Right] The CDF lepton charge asymmetry is compared with predictions
 with CTEQ6M and MRST02 PDFs using a NLO RESBOS calculation.}
  \label{fig:du}
\end{figure}

The ratio of $d$ and $u$ quarks at high $x$ primary comes from the measurements
of $F_2$(deuterium) and $F_2$(proton). Because of a large uncertainty of nuclear binding effect on deuterium target, this ratio is poorly known. Figure~\ref{fig:du}[left] shows the NMC
$F_2^d$/$F_2^p$ with and without nuclear correction~\cite{kuhlman}. 
With a nuclear correction,
the NMC data favors 0.2 for $d/u$ as $x \rightarrow 1$, which is of theoretical
interest for nuclear physics community. However, the size of nuclear binding correction
is still controversial.
S.Kuhn~\cite{kuhn} presented dedicated JLab efforts to study $d/u$ at high $x$.
Their programs are 
to study the effect of nuclear binding on neutron structure, and to measure the structure functions
of a free neutron by detecting a slow spectator proton. Information on $d/u$ can be also
extracted from $W$ production data at the Tevatron.
The CDF collaboration~\cite{chung} measured the forward-backward charge asymmetry of electrons from $W$ boson
decays. In order to get a better $d/u$ sensitivity on higher $x$ region, they  have looked 
at a higher electron $E_T$ region. Fig~\ref{fig:du}[right] shows comparisons
with the NLO RESBOS predictions using CTEQ6M and MRST 2001 PDFs. 
At high $\eta$, the CDF data tends to favor higher
$d/u$ value at high $x$. Thus, it would be interesting to compare with the PDFs which
was extracted, assuming a large nuclear correction on the deuterium target. They expect to 
have a big improvement on this measurement by reconstructing $W$ rapidity directly.

\begin{figure}
\includegraphics[width=.5\columnwidth,height=.4\textheight]{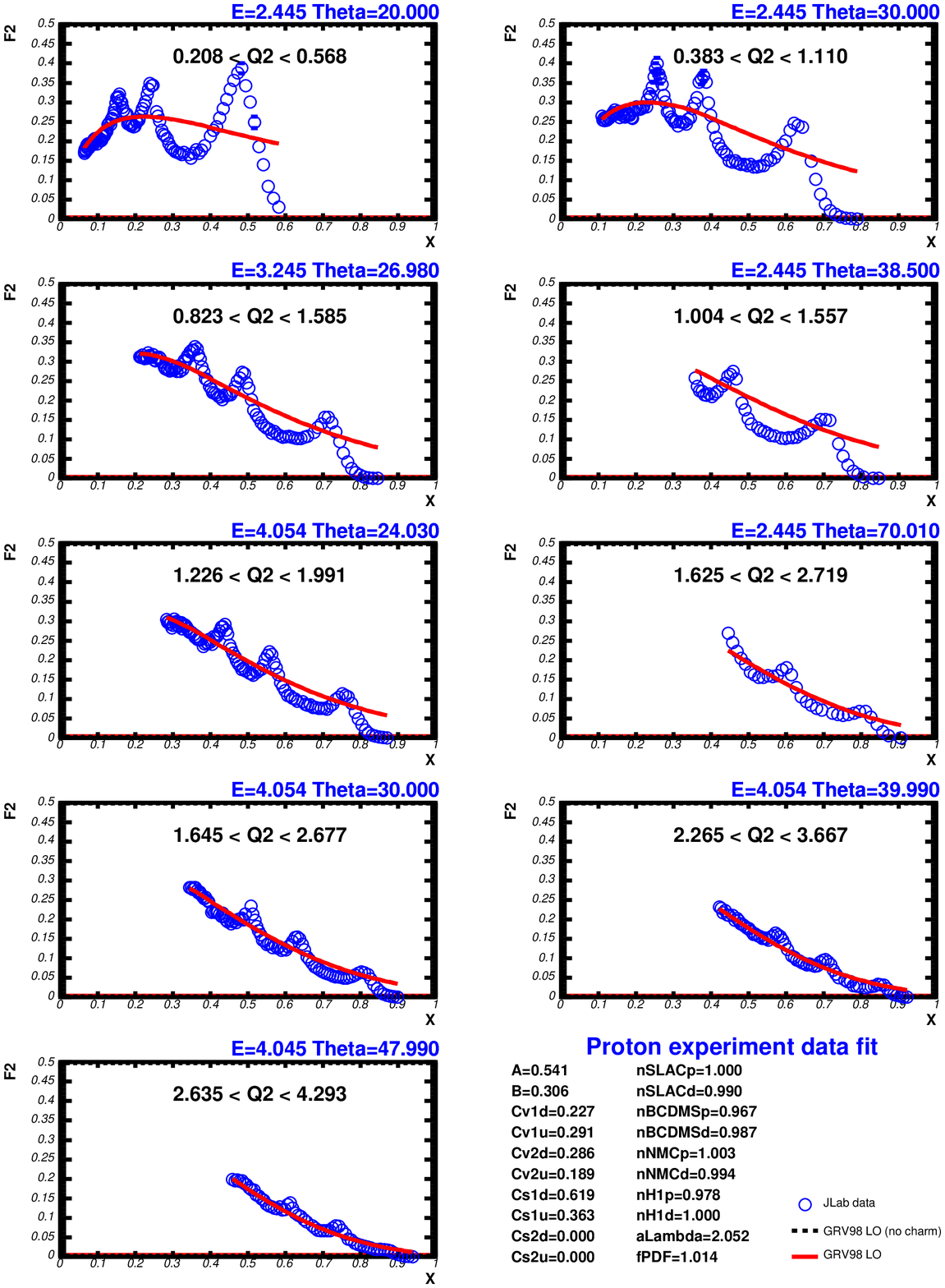}
\includegraphics[width=.5\columnwidth,height=.4\textheight]{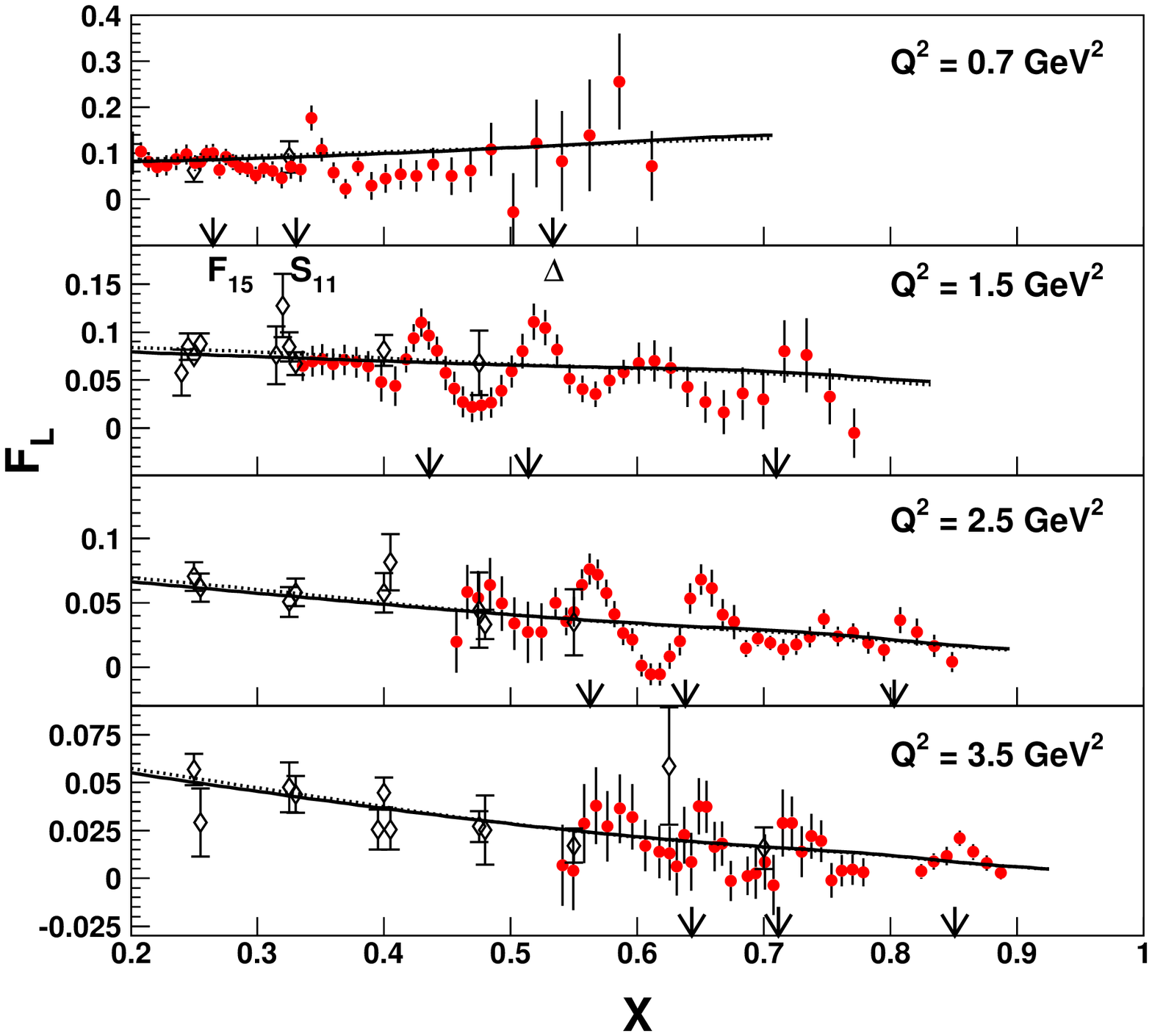}
\caption{ Comparisons of the predictions of Bodek-Yang 
model to resonance electro-production data on proton:
JLab $F_2$ proton [left], and $F_L$ proton[right].}
\label{fig:bodek}
\end{figure}

At the workshop, one of the hot subjects was the phenomenon of a parton-hadron duality which
states that the average behavior of the nucleon resonances follows the DIS scaling limit curve.
JLab has very precise data at high $x$ and low $Q^2$ region (where a resonance production occurs).
Besides many theoretical issues discussed by S. Liuti~\cite{liuti},
C. Keppel and I. Niculescu~\cite{jlab} demonstrated that the duality holds 
for $F_2$ proton, EMC effect, and even spin structure functions in the region of $Q^2>0.5$ GeV$^2$.
Issues are whether non-perturbative power corrections between DIS and resonance region is same,
and DGLAP evolution \& factorization works in the resonance region too. S. Liuti's studies suggest
that the size of the higher twist effect may not be same.
However, A. Bodek~\cite{bodek} showed that all DIS $F_2$ data 
and JLab's resonance data
are well described by his Bodek-Yang leading-order model,
as shown in Figure~\ref{fig:bodek}. This implies that there is not much 
difference in the power corrections between two regions.
This model uses a new scaling
variable $\xi_w$ and $Q^2$ dependent $K$ factors to all PDFs
to describe both pQCD and non-perturbative QCD regions very smoothly.

A community of neutrino oscillation physics have started to pay attention
on this non-perturbative QCD region. For precise measurements
of mass splitting and mixing angles, neutrino oscillation
experiments (MINOS, NO$\nu$A, and T2K) need to have a good understanding
of neutrino cross section at low energy. This point was well presented
by H. Gallagher~\cite{osci}. Certainly this would be a place 
where DIS, nuclear physics, and neutrino physics communities 
need to make a coherent effort.

\begin{figure}
\includegraphics[width=.5\columnwidth,height=.35\textheight]{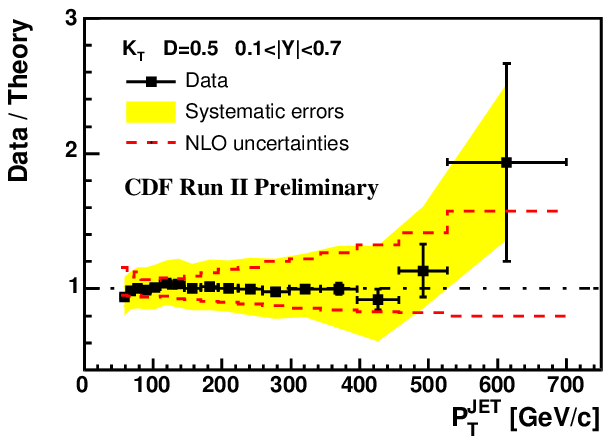}
\includegraphics[width=.5\columnwidth,height=.37\textheight]{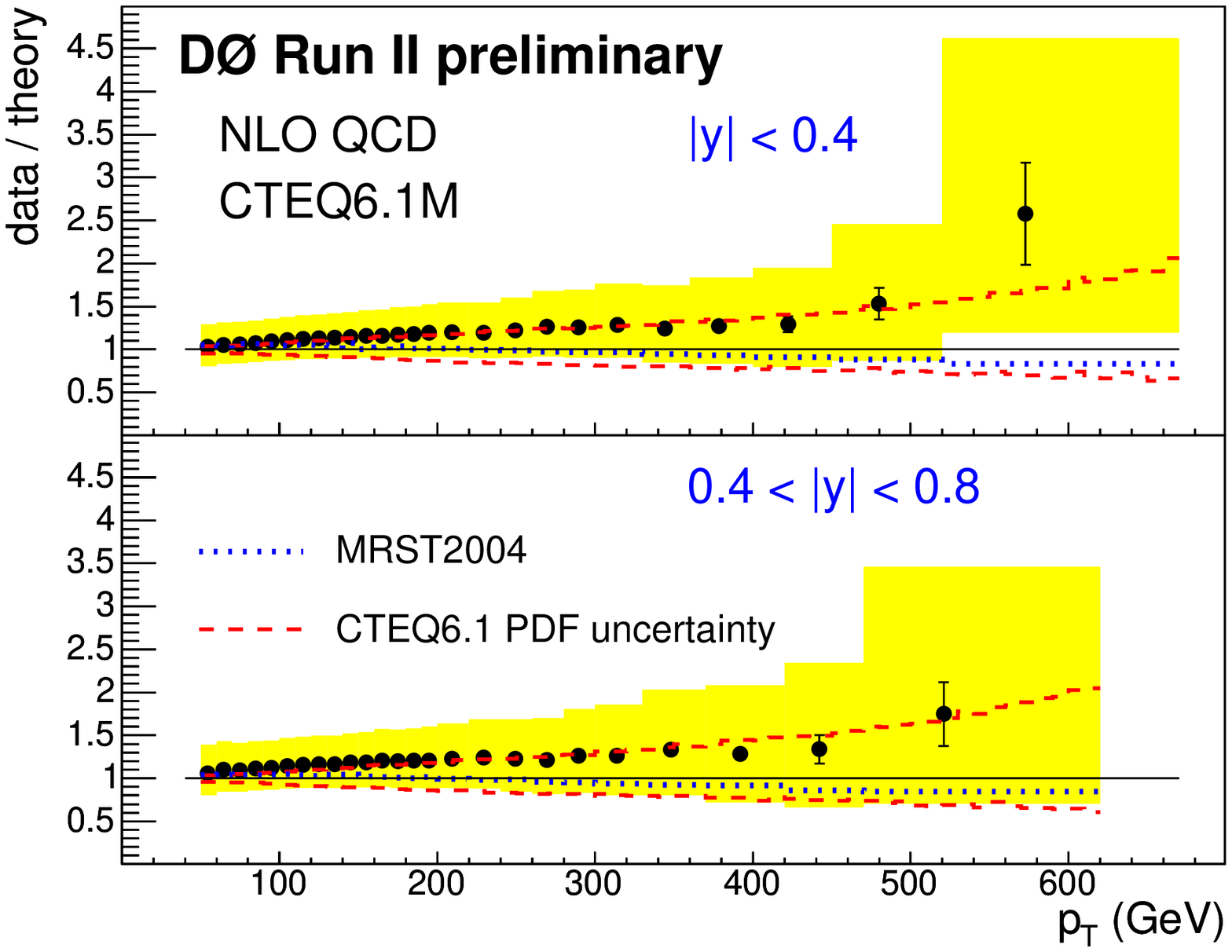}
\caption{ The ratios of the inclusive jet cross section data and the NLO predictions:
CDF data using $K_T$ algorithm [left] and D0 data using cone algorithm[right].}
\label{fig:jet}
\end{figure}

At the end of the workshop, PDF uncertainties at the Tevatron 
and impact on various measurements are presented by F. Chlebana,
and A. Harel~\cite{jet}. Chelbana discussed many ideas to constrain
the PDFs using the Tevatron data. 
Figure~\ref{fig:jet} shows the latest status of the Tevatron jet data
with the NLO predictions where there is no observed discrepancy
at high $E_T$ region. These measurements are currently dominated
by the jet energy scale uncertainties. In future, it would
be important to separate PDF effects from any new physics signal.
F. Chlebana He also pointed out that it is crucial to measure
the size of heavy flavor quarks densities for Higgs and Top physics.

\section{Progress in the determination of parton distribution functions}

The structure functions discussed in the previous section can be expressed in
terms of the parton distribution functions (PDFs) of the proton.  The structure
function, $F_2 (x,Q^2)$ is proportional to the sum of the quark and antiquark 
PDFs.  At low $x$, $F_2$ is therefore sensitive to the sea quark distributions
and hence is indirectly sensitive to the gluon density.  The longitudinal
structure function, $F_L$, is directly sensitive to the gluon density.
The parity-violating structure function, $xF_3 (x,Q^2)$, is proportional to 
the difference between the quark and antiquark PDFs, making it sensitive to 
the valence quark distributions.

The PDFs of the proton may be extracted by fitting, among other things,
the HERA structure function data.  These fits have been performed by a 
number of different groups, as well as the experimental collaborations
themselves.  It is crucial that the best possible understanding of the 
proton PDFs is achieved, given their central role in predictions for 
other processes, for example, at the LHC.

The traditional method of determining the PDFs of the proton relies on
assuming an $x$-dependence for each of the different PDFs at some starting 
scale $Q_{0}^{2}$ and then using the Dokshitzer-Gribov-Lipatov-Altarelli-Parisi
(DGLAP) evolution equations~\cite{dglap} to model the $Q^2$ dependence of
the PDFs.  This approach is used by both the CTEQ Collaboration~\cite{cteq} 
and Martin et al.\ (MRST)~\cite{mrst}; both groups presented progress reports 
at this workshop.  This approach has also been adopted by the ZEUS 
Collaboration~\cite{zfit}, who also presented results of their latest fit at 
this workshop.

A number of issues have been addressed by both the CTEQ and MRST groups 
recently. In particular, the compatibility of datasets and the stability of 
the fit results have been studied.

\begin{figure}
  \includegraphics[height=.3\textheight]{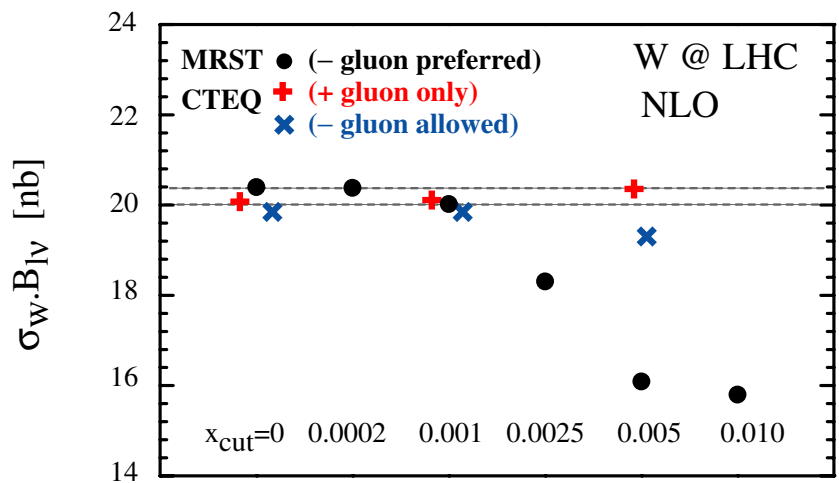}
  \caption{Predictions for the $W^{\pm}$ production cross section at the LHC from the CTEQ and MRST Collaborations.  The cross sections are plotted as a function of the lower $x$ limit applied to the data used to extract the PDFs.  The CTEQ Collaboration have considered two different scenarios:  one in which the 
gluon is forced to be positive-definite and the other in which the gluon is
left free.  Both are indicated by crosses.  The MRST predictions are indicated
by the dots and here the gluon is left free.}
  \label{fig:cteq}
\end{figure}

Both groups have performed studies of fit stability~\cite{cstab,mstab},
by studying the impact of restricting the fits to data at higher $x$.  The 
studies were performed by looking at the next-to-leading order (NLO) 
$W^{\pm}$ production cross section predictions from fits with different lower 
$x$ limits.  The results of both studies are shown in figure~\ref{fig:cteq}.  
The CTEQ group conclude from their studies that the fits are stable.  The MRST
group conclude that the uncertainties increase significantly as the lower $x$ 
limit is tightened and that next-to-next-to-leading-order (NNLO) is inherently
more stable and these fits should become the standard in the future.

The MRST group also presented the results of other studies, in particular
the inclusion of electroweak corrections and QED effects~\cite{mqed}.  The 
latter, in particular, have a negligible effect on the PDFs themselves, as 
expected.  However, the inclusion of QED effects does lead to a small isospin 
violation, which significantly improves the predictions for prompt photon 
production at HERA.

The ZEUS Collaboration also presented their latest determination of the proton
PDFs using only ZEUS data~\cite{zfpap}.  In comparison to their previous 
fits, ZEUS jet cross section data has been included, which is directly 
sensitive to the gluon density of the proton and benefit from small 
experimental and theoretical uncertainties.  Both NC DIS jet data and 
direct-enriched dijet photoproduction data, in which the photon behaves as
a point-like object, have been included, significantly improving the
uncertainty on the gluon PDF in the range $0.01 < x < 0.4$.  If a simultaneous
fit of both the PDFs and $\alpha_s$ is performed, the result for $\alpha_s$
is very precise and in agreement with the world average.

Several presentations were also made at this workshop, in which alternative
approaches to PDF determination were explained.  One such presentation was
made by the NNPDF Collaboration~\cite{nnpdf}, who are developing a neural 
network approach to PDF fitting.  This approach avoids some of the 
shortcomings of the standard method, such as avoiding any potential bias from 
the choice of functional form for the PDFs.  It should also lead to a better 
estimation of the PDF uncertainties.  So far the structure functions have been
successfully determined using this approach, but work is still in progress to 
successfully determine the PDFs using this method.

The estimation and reduction of PDF uncertainties was a key theme at this
workshop.  One presentation made at this workshop looked at the possibility
of averaging the $F_2$ data from the H1 and ZEUS experiments, prior to 
including it in any global PDF fit~\cite{glazov}.  This has 
advantages when it comes to the handling of the systematic uncertainties;
it also provides a model-independent method of checking the consistency
of the data from the two experiments.  It has been found that several
contributions to the systematic uncertainties from each experiment are 
reduced; the experiments are effectively constraining each other.  This is
an interesting approach which, it is hoped, will be pursued further by the
two Collaborations.

Another presentation made at this workshop looked at the impact of future HERA
data on the PDF uncertainties~\cite{gwen}.  This study was performed using the
ZEUS PDF fit as a basis.  A number of different scenarios were considered, 
including simply the expected increase in the amount of luminosity, as well
as the inclusion of new cross section measurements which have been optimized 
to constrain the PDFs as tightly as possible.  These improvements would lead 
to significant improvements in the valence quark distributions, as well as in 
the high $x$ sea quark and gluon distributions.  Other scenarios which were
also considered are the inclusion of precision measurements of $F_L$ from
HERA data (low-energy proton running) and the possibility of $e-D$ running
to constrain the sea quark asymmetries.

\section{Toward QCD precision tests}

New HERA data on unpolarized DIS structure functions, combined
with the present world data, allow to reduce the experimental
error on the strong coupling constant, $\alpha_s(M_Z^2)$, the
fundamental constant of QCD and strong interaction, to the level of 
$1$\%.  On the theoretial side, the next-to-leading order (NLO)
analyses have limitations due to scale variations being present which 
allow no better than $5$\% accuracy in the determination of $\alpha_s$ 
~\cite{guffanti}.  In order to match the experimental accuracy, it was
stressed ~\cite{guffanti} that analyses of DIS structure functions need
to be carried out at the NNLO level.  With the recent 
computation of the 3--loop anomalous dimensions ~\cite{Moch:2004pa}, 
a complete NNLO study of DIS structure functions is now possible.   
A full NNLO analysis of unpolarized DIS structure functions aiming to
obtain a high accuracy determination of $\alpha_s$ was presented at
the workshop ~\cite{guffanti}.  It was pointed out that a combination
of standard NNLO QCD analysis and fits based on factorization
scheme-invariant evolution of DIS structure functions will provide a
valuable tool in high-precision analyses aiming at $1$\% accuracy in
the determination of $\alpha_s$ ~\cite{guffanti}.  The factorization
scheme-invariant evolution of DIS structure functions can be
implemented to different pair of structure functions, such as $F_2$
and $F_L$ or $F_2$ and its $t \propto \ln(\alpha_s(Q))$ derivative.  
In this approach, $\alpha_s$ is determined by performing an
one dimensional fit between the evolution of DIS structure functions
and the data.  Work is still ongoing. A full NNLO accuracy evolution
for $F_2$ and $\partial F_2/\partial t$ have been completely
implemented for massless flavors.  Inclusion of heavy flavors and the
fit to the data, and the one parameter fit to determine $\Lambda_{\rm
  QCD}$ are on the way.  One interesting result is that comparing the
behavior of slopes of $\partial F_2/\partial t$ to the slopes
extracted experimentally points toward a positive gluon density in
small-$x$ region ~\cite{guffanti}, while NLO global analysis of PDFs
points to a negative gluon density in low-$x$ and low $Q^2$ region
~\cite{mrst}.  

An effort to develop a next generation of event generators was
reported at the workshop ~\cite{zu}.  The effort was aiming to set up
a systematic scheme for developing event generators that are
consistent to QCD factorization of differential cross sections up to
NLO accuracy.  It was argued that in order to achieve this accuracy,
one has to use unintegrated PDFs to replace the parton shower in the
LO event generators.  The basic rules have been established for a DIS
event generator, but, there are still works to be done ~\cite{zu}.

At hadron colliders, it is the PDFs that determine partonic flux of
hard collisions.  Full discovery potentials of the LHC and precision
tests of QCD are sensitive to PDFs at large $x$. In the form of QCD
factorization, extraction of PDFs
depends on short-distance dynamics and perturbatively calculated
coefficient functions.  The coefficient functions often have high
powers of logarithms like $\ln(1-x)$ and $\ln(1/x)$, which become
large as $x$ near 1 and 0.  Resummation of these large logarithms is
necessary for observables dominated by those kinematic regions.  
A presentation made at this workshop looked at the effect of large-$x$
resummation on the extraction of PDFs ~\cite{corcella}.  Large-$x$
resummation was performed for coefficient functions of DIS structure
functions in massless approximation as well as in an approach that
includes heavy quark-mass effects.  After performing fits to the fixed
target DIS data from NuTeV, BCDMS and NMC collaborations, using NLO
and NLL-resummed coefficient functions, it was found that the
resummation has a visible impact on the extraction of quark
distributions at large $x$, and was stressed that large-$x$ partonic 
resummation is needed whenever a high precision is required for cross
sections evaluated near partonic threshold ~\cite{corcella}.

A precise knowledge of PDFs, in particular, gluon distribution at
$x \sim 0.001-0.01$ are vital for understanding almost all standard
production processes at the LHC.  When we move away from zero
rapidity, much smaller $x$ partons, as small as $10^{-5}$, are
required for some observables.  
Although perturbative QCD has been very successful in
interpreting data on scaling violation of PDFs in terms of DGLAP
evolution in $Q^2$ ~\cite{cteq,mrst}, the extrapolation of PDFs to
smaller $x$ has not been very consistent with the BFKL evolution in
$x$ (or in energy) ~\cite{bfkl}.  Although the strong rise of proton
structure function $F_2$ with energy, observed at HERA, can be
well described by a simple (3 parameters) LO BFKL fit ~\cite{LO_BFKL},
a much too small effective $\alpha_s < 0.1$ is needed while the world
average is $\alpha_s\sim 0.2$ for HERA kinematics.   
A phenomenological study of confronting NLO BFKL with new HERA data on
$F_2$ structure function was presented at the workshop ~\cite{royon}.  A
big discrepancy between theory and data, especially at low $Q^2$, was
clearly evident ~\cite{royon}.  A further study is needed although more
progresses have been made recently ~\cite{forte}.

\section{Low-$x$ physics: parton evolution and saturation}

One of the challenging problems in QCD is to understand the behavior of
hadronic cross sections in high energy limit.  Experimental data on
the total cross section show a slow but distinct rise with collision
energy $\sqrt{s}$.  This rise could be parametrized by a power of $s$,
$\sigma(s) \sim s^{0.08}$, which is consistent with an exchange of
soft Pomerons ~\cite{SP}. 
On the other hand, after resumming leading powers of $\ln(s)$
contributions, perturbative QCD calculation, in the form of BFKL
evolution in energy (or in $x$), predicts a much stronger rise with a
much large power of $s$ ~\cite{bfkl}.  As $\sqrt{s}\rightarrow \infty$ (or
$x\rightarrow 0$), the power-like rise is not compatible with the
unitarity of the S-matrix in the high energy limit, or in
contradiction with the Froissart bound ~\cite{FB}, which allows at most
a logarithmic increase with collision energy.

BFKL equation is a linear evolution equation and predicts a large
number of low-$x$ partons due to the strength of soft gluon radiation
in QCD.  On the other hand, the large number of soft partons generated
by parton radiation are likely to interact and recombine. Parton
recombination introduces non-linear terms into the BFKL equation, 
slows down the small-$x$ evolution, and removes the apparent
violation of the unitarity.  When parton
recombination is strong enough to balance parton radiation, PDFs
saturate as $x\rightarrow 0$ ~\cite{recombination}.  
The state of saturated partons is sometime referred as the Color Glass
Condensate (CGC) ~\cite{mv,jimwlk,CGC}.   
A lot of work, both theoretical and experimental, have been done and
many progresses have been made in recent years to understand this
saturation phenomenon, especially, in a nuclear environment because of
an $A^{1/3}$ length enhancement in parton density at a given impact
parameter.  Two sessions at this workshop were devoted to
the presentations related to this novel phenomenon.

A simple modification to the BFKL equation is Balitsky-Kovchekov (BK)
equation ~\cite{bk}, which adds a quadratic term to the BFKL equation.
The BK equation is a non-linear integro-differential equation for
unintegrated PDFs.  Its non-linearity leads to many interesting
features that could be seen in high energy reactions.  It was
shown ~\cite{MP} that the BK equation is in the equivalence class
of the Fisher--Kolmogorov--Petrovsky--Piscounov (FKPP)  non-linear
partial differential equation, which has so-called traveling wave
solutions.  The similarity leads to an interesting point of view
that high energy QCD is equivalent to a reaction--diffusion system
~\cite{enberg}.  A detailed numerical studies of the mean field
approximation to the BK equation was presented at the workshop
~\cite{enberg}.  It was demonstrated that the numerical solutions of
the BK equation does show features of traveling wave solutions.
It was also confirmed that the influence of the initial condition
disappears for large $Y\sim \ln(1/x)$, so that a universal propagation
speed is approached, which should help establish statistical
interpretations of the phenomena observed in QCD scattering at high
energy. 

A study of discrete version of the BK equation was presented at this
workshop ~\cite{tuchin}.  By noting that the number of gluons in the
hadron wave functions is discrete, and their formation in the chain of
small $x$ evolution occurs in the discrete intervals of $\ln(1/x)$, a
discrete version of BK equation was formulated ~\cite{tuchin}.  It was
found that numerical solutions of the discrete BK equation behave
chaotically in the phenomenologically interesting kinematic region.
It was concluded ~\cite{tuchin} that the evolution of the scattering
amplitude at high energies in the saturation region might be chaotic,
while the scattering amplitude in the normal perturbative region 
is not affected by the discretization. 
Although the model used in the numerical calculations 
neglected the diffusion in transverse momentum, stochasticity of
gluon emission and the dynamical fluctuations beyond the mean field
approximation, it was hoped that at least some of the features of
discrete quantum evolution at small $x$ will survive a more realistic
treatment.  The chaotic features of small $x$ evolution   
open a new intriguing prospective on the studies of hadron and nuclear
interactions at high energies. 

Recent developments of CGC theory were reported at the workshop
~\cite{stasto,kovner}. The CGC theory  is an effective theory of the
strong interactions at very high energies ~\cite{CGC}.   
The basic equation of CGC theory is the JIMWLK equation ~\cite{jimwlk},
which governs the evolution of a weight function of a color medium (or
a hadron target) with rapidity.  The evolution kernel is often
referred as the JIMWLK Hamiltonian.  
The weight function is needed for calculating
physical scattering amplitude when it is averaged over the medium's
color charge. In the large $N_c$ limit and in the dipole scattering
picture, the JIMWLK equation reduces to the closed and relatively simple
BK equation. In a language of Feynman diagrams, the JIMWLK equation 
includes both BFKL ladder diagrams and the fan diagrams of triple
ladder interactions, and it naturally describes the physics of
scattering on a dense medium (or a target) with multiple scattering
corrections.  It naturally interprets the geometric scaling observed
in the data ~\cite{CGC}.  However, the JIMWLK equation does not include
Pomeron (or ladder) splittings or all the Pomeron loops.
Modifications and improvements to the JIMWLK equation were proposed
~\cite{stasto,kovner}.  In addition, a similar evolution equation was
derived for a dilute target ~\cite{kovner}, while the JIMWLK equation
is suitable for scatterings on a dense target.  A striking result is
that the evolution kernels of these two equations are apparently dual
to each other ~\cite{stasto,kovner}.  The selfduality of the kernel 
is somewhat similar (although different in detail) to the duality
symmetry, $p\rightarrow x$, $x\rightarrow -p$ in the Hamiltonian of a
harmonic oscillator ~\cite{kovner}.
 
Relativistic Heavy Ion Collider (RHIC) is a unique place to test the
theory of CGC because of the high density of partons involved.
In order to probe small $x$ partons and the phenomena of CGC at RHIC,
one has to go to extremely forward and backward region in rapidity
because of the relatively low colliding energy.  Three major
experimental collaborations, BRAHMS, PHENIX, and STAR, at RHIC carried
out the effort and presented their early results at the workshop
~\cite{brahms,phenix,star}. 

\begin{figure}
\includegraphics[width=1.0\columnwidth]{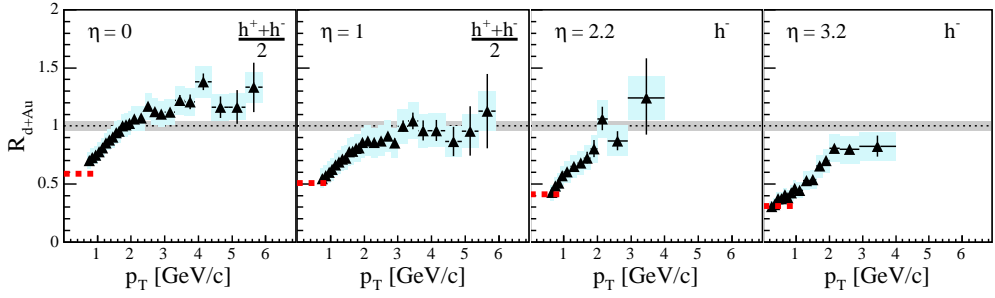}
\caption{Nuclear modification factor for charged hadrons at pseudo-rapidities 
$\eta=0,1.0,2.2,3.2$. Statistical errors are shown with error bars.
Systematic errors are shown with shaded boxes with widths set by the
bin sizes.  The shaded band around unity indicates the estimated error
on the normalization to $\langle N_{coll}\rangle$.  Dashed lines at
$p_T < 1$~GeV/c show the normalized charged particle density ratio
$\frac{1}{\langle N_{coll}\rangle}
 \frac{dN/d\eta(dAu)}{dN/d\eta(pp)}$}
\label{fig:brahms}
\end{figure}

Rapidity dependence of high-$p_T$ particle suppression was measured
in d-Au collisions at $\sqrt{s_{NN}}=200$~GeV by BRAHMS Collaboration
and presented at the workshop ~\cite{brahms}.  
The data collected from d-Au collisions at RHIC is compared to p-p
in Figure~\ref{fig:brahms} using the nuclear modification factor defined as 
\begin{equation}
R_{\rm d-Au} = \frac{1}{N_{coll}}\
    \frac{\frac{dN^{\rm dAu}}{dp_Td\eta}}
         {\frac{dN^{\rm pp}}{dp_Td\eta}}
\label{Rda}
\end{equation}
where $N_{coll}$ is the number of binary collisions estimated to be 
$7.2\pm 0.6$ for minimum biased d+Au collisions.    
Nuclear modification factors for charged hadrons at pseudo-rapidities 
$\eta=0,1.0,2.2$ and 3.2 were shown as a function of hadron transverse
momentum $p_T$.  At the central region, or zero rapidity, data confirm
the Cronin type enhancement in large $p_T$ region.  However, as the
pseudo-rapidity increases, the enhancement vanishes, and the
modification factor is less than the unity for entire measured $p_T$
region.  The forward region is, the measurement probes target partons
with smaller $x$.  The observed rapidity dependence of the
suppression, which increases with rapidity, 
fits naturally into the picture of CGC ~\cite{kovchegov}, 
and can be also interpreted by the recombination model of
hadronization ~\cite{fries}.  In addition, the suppression is
consistent with the perturbative QCD calculation based on resummation
of coherent multiple scattering ~\cite{vitev}.

\begin{figure}
\includegraphics[width=.8\columnwidth]{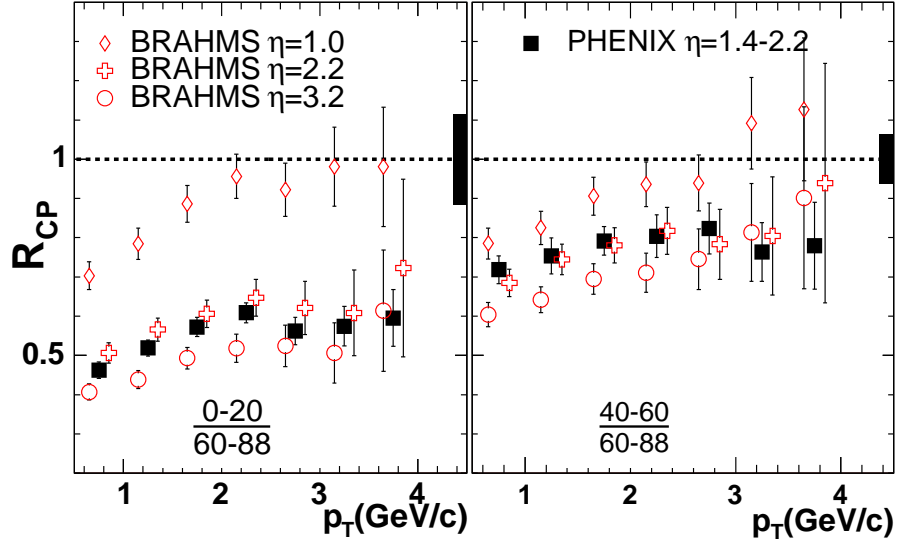}
\caption{PHENIX $R_{cp}$ as a function of $p_{T}$ at forward rapidities shown
as the average of the two methods.  Note that the BRAHMS results are
for negative hadrons at $\eta=2.2,3.2$ and their centrality ranges
($0-20\%/60-80\%$ and $30-50\%/60-80\%$) are somewhat different from
ours.}
\label{fig:phenix}
\end{figure}

PHENIX Collaboration reported its measurement of charged hadron
production in the same d-Au collisions at RHIC ~\cite{phenix}.  
It covers pseudo-rapidities from -2.0 to -1.4 and 1.4 to 2.2 with
the forward coverage overlaps with some of BRAHMS measurements.
PHENIX also observes a suppression in hadron yields in d-Au collision
relative to binary collision.  The data was presented in terms of   
a different nuclear modification factor, $R_{cp}$, which is defined as
the ratio of the particle yield in central collisions to the particle
yield in peripheral collisions, each normalized by the averaged number
of binary collisions $N_{coll}$
\begin{equation}
R_{\rm cp} = 
\frac{\left. \frac{dN^{\rm central}}{dp_Td\eta} 
      \right/ N_{coll}^{\rm central} }
     {\left. \frac{dN^{\rm peripheral}}{dp_Td\eta}  
      \right/ N_{coll}^{\rm peripheral} } .
\label{Rcp}
\end{equation}
As shown in Figure~\ref{fig:phenix}, PHENIX data are consistent with BRAHMS data, and are in qualitative
agreement with theoretical expectation.  Quantitatively, the ratio
$R_{cp}$ for the most central over the most peripheral collisions is
more suppressed than theoretical calculations.

\begin{figure}
\includegraphics[width=.5\columnwidth]{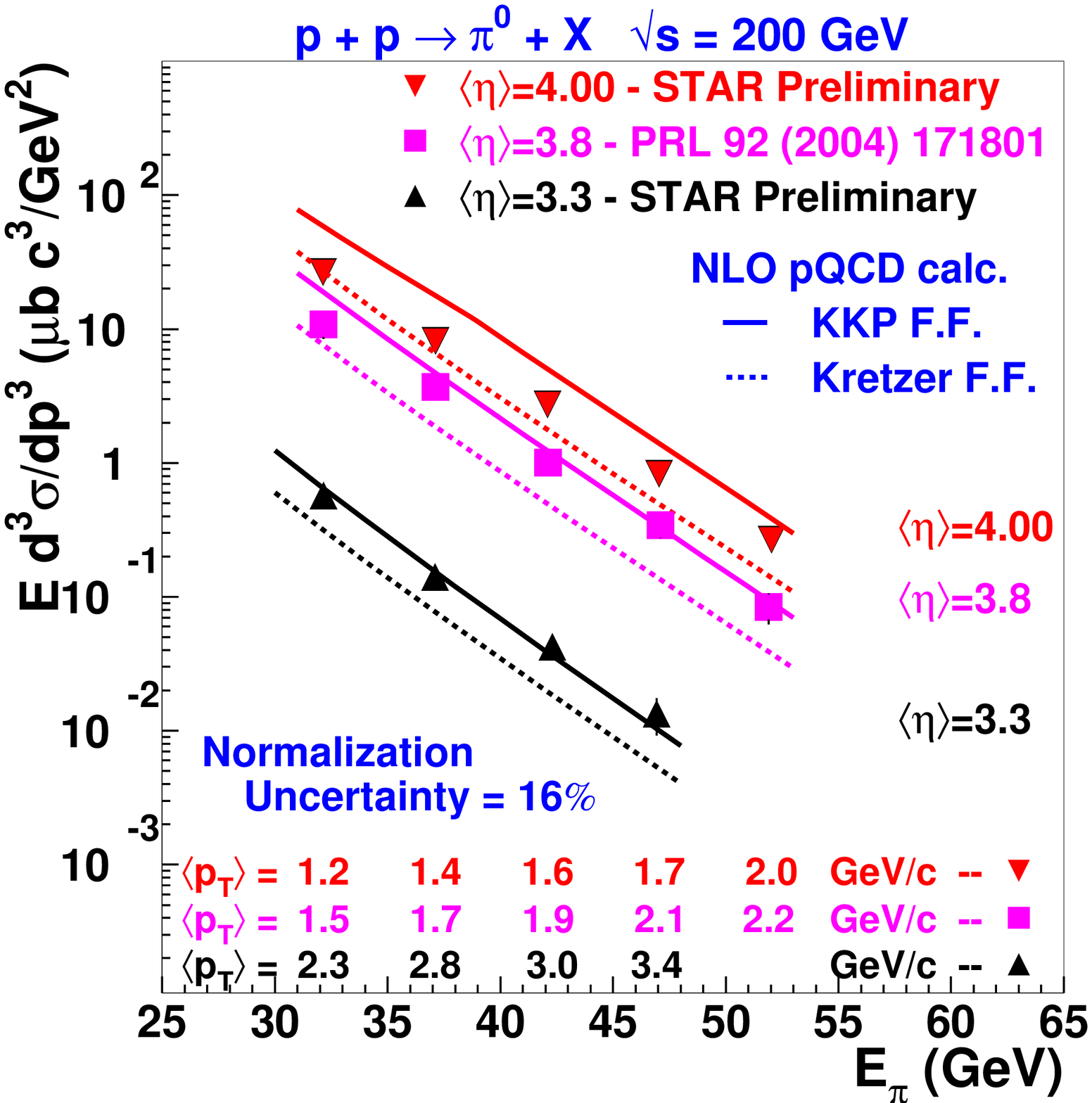}
\includegraphics[width=.5\columnwidth]{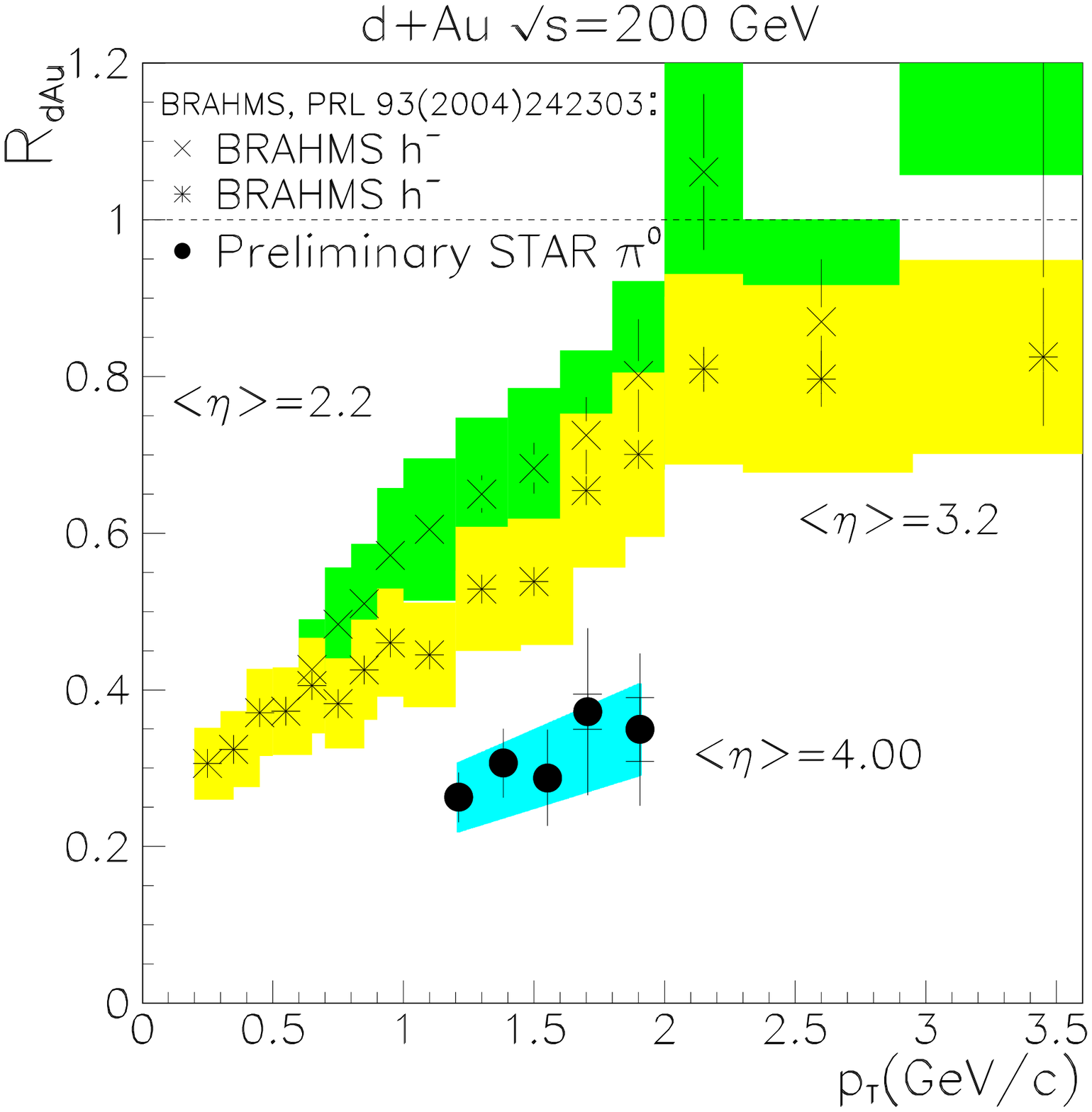}
\caption{Inclusive $\pi^0$ yield for p+p [left] and d+Au collisions 
normalized by p+p [right].
The pion energy ($E_\pi$) is correlated with the transverse 
momentum ($p_T$), as the FPD was at fixed values of 
pseudo-rapidity ($\eta$).
The inner error bars are statistical, while the outer combine 
these with the $E_\pi$- ($p_T$-) dependent systematic errors,
and are often smaller than the points.
The curves (left) are NLO pQCD calculations evaluated at fixed 
$\eta$, using different fragmentation functions.
The x's and stars (right) are BRAHMS data for $h^-$ 
production at smaller $\eta$.}
\label{fig:star}
\end{figure}

Measurements of the inclusive yields of $\pi^0$ mesons in p-p and d-Au
collisions at RHIC were presented by STAR Collaboration ~\cite{star}.
With a forward $\pi^0$ detector installed at the Solenoidal Tracker at
RHIC (STAR), it can detect high energy $\pi^0$ mesons with
pseudo-rapidity as large as $3.3 < \eta < 4.1$ ~\cite{star}.  The
inclusive yield in p-p collisions at $\sqrt{s}=200$~GeV are consistent
with NLO pQCD calculations.  The nuclear modification factor,
$R_{dAu}$ in Figure~\ref{fig:star}, 
shows a strong suppression at the large pseudo-rapidity.  It
was argued that the d-Au yield is consistent with a model calculation
treating the Au nucleus as a CGC for forward particle production
~\cite{star}.  Comparisons with other production models will be
interesting to perform.  Additional measurements with different
final-state particles and at different centralities will help
elucidate the cause of the observed strong suppression, which covers a
broad range of nuclear gluon momentum fraction with a peak value
$x\sim 0.02$ ~\cite{star}.     

Two theory talks were presented at the workshop to specifically
address the strong suppression observed in the forward region of d-Au
collisions at RHIC ~\cite{kovchegov,fries}.  Two completely different
pictures were presented on how a leading hadron was produced in the
d-Au collisions at RHIC energies.
In one approach ~\cite{kovchegov}, single hadron production was assumed
to be proportional to gluon production. Under this approximation,
the nuclear modification factor $R_{dAu}$ is the same for both hadron
and gluon production.  The gluon production in d-Au collisions was
calculated in the framework of CGC physics ~\cite{kovchegov}. 
In the other approach ~\cite{fries}, a single hadron was produced via
recombination of partons available during the collisions.  It was
argued that p-p, p(d)-Au, and Au-Au collisions produce different
shapes of parton spectra.  Recombination of partons with different
spectra naturally leads to different hadron distribution and a
nontrivial nuclear modification factor ~\cite{fries}. 
A striking fact is that both of these approaches provided 
calculations that are consistent with the observed data.

\section{Nuclear structure functions and nuclear PDFs}

It was observed about two decades ago that DIS structure functions of
nuclei differ from simple sum of those in the free nucleon ~\cite{EMC}.
As a result, PDFs of a nucleus of atomic weight $A$ also differ from
those in the free proton, $f_i^A(x,Q^2) \ne f_i(x,Q^2)$.  In order to
understand the overwhelming data from the RHIC and make predictions
for the heavy ion programs at the future facilities, like the LHC and  
Electron Ion Collider (EIC), we need precise information of nuclear
PDFs (nPDFs), in particular, at small $x(<0.1)$.

A brief overview of the global DGLAP analyses of nPDFs was presented
at the workshop ~\cite{kolhinen}.  The nPDFs are defined in terms of
the same operators that define the free nucleon PDFs with the free
nucleon state replaced by a nuclear state.  Therefore, nPDFs and free
nucleon PDFs should share the same DGLAP evolution equations, and  
only difference between nPDFs of different nuclei and free nucleon
PDFs is the input distributions to DGLAP equations at a scale $Q_0^2$.
Once a set of the nonperturbative input distributions are chosen,
DGLAP evolution equations predict nPDFs at a larger momentum scale
$Q^2$.  There are typically two approaches to choose the input
distributions: calculated by the nuclear models and determined by
fit to the data ~\cite{kolhinen}.  The second approach shares the same
procedures as that used in the determination of PDFs, and often
referred as the global analyses of nPDFs.  There are three groups who
have been carrying out the global analyses and reanalyses of nPDFs: 
Eskola {\it et al.} (usually called as {\it EKS98})
~\cite{Eskola:1998iy,Eskola:1998df,EKS05}, 
Hirai {\it et al.} ~\cite{Hirai:2001np,Hirai:2004wq}, and
de~Florian and Sassot ~\cite{deFlorian:2003qf}. 
The first two groups use LO DGLAP evolution while the third uses NLO
evolution.  All analyses, only DIS and Drell-Yan data on nuclear
targets were used in the fits.  Because of the large error in nuclear
data and the lack of direct information on gluon initiated processes,
all fits have reasonable constrains and consistencies on quark
distributions, but, not on gluon distributions ~\cite{kolhinen}.  
  
A hard probe often refers to a scattering process with a large
momentum exchange $q^\mu$ whose invariant mass $Q\equiv \sqrt{|q^2|}
\gg \Lambda_{\rm QCD}$, and it can probe a distance scale much
smaller than size of a nucleon at rest, $1/Q\ll $~fm.
However, when an active parton's momentum fraction $x < x_c \sim 0.1$,
a hard probe might interact with more than one partons of the nucleon
coherently ~\cite{Qiu:2002mh}.  When $x \ll x_c$, the hard probe can  
cover a whole Lorentz contracted nucleus and interact with partons
from different nucleons.  Although such coherent multi-parton
interactions are power suppressed by hard scales of the scattering, 
they are enhanced by the nuclear size and could be one of the
important sources of nuclear dependence observed in high energy
nuclear collisions.  A presentation made at this workshop looked at
the impact of coherent multiple scattering in DIS on nuclear targets and
leading particle production in p(d)-Au collisions ~\cite{vitev}.
An all power resummation of nuclear enhanced power corrections to DIS 
structure functions on nuclear targets was achieved.  The calculated
results for the Bjorken $x$-, $Q^2$- and $A$-dependence of nuclear
shadowing in $F_2^A(x,Q^2)$ and the nuclear modifications to
$F_L^A(x,Q^2)$ are consistent with the existing data ~\cite{vitev}.
Predictions were made for the dynamical shadowing from final state
interactions in $\nu + A$ reactions for sea and valence quarks in the
structure functions $F_2^A(x,Q^2)$ and $x F_3^A(x,Q^2)$,
respectively.  In addition, calculations for the centrality and
rapidity dependent nuclear suppression of single and double inclusive
hadron production at moderate transverse momenta in $p+A$ collisions
were presented and consistent with the RHIC data ~\cite{vitev}.

In the Gribov-Glauber picture, nuclear shadowing and antishadowing
observed in nuclear structure functions are due to the destructive and
constructive interference of amplitudes arising from the
multiple-scattering of quarks in the nucleus, respectively.
A calculation of shadowing and antishadowing of nuclear structure
functions in the Gribov-Glauber picture were presented at the workshop
~\cite{brodsky}.  The coherence of multi-step nuclear processes
leads to shadowing and antishadowing of the electromagnetic
nuclear structure functions in agreement with the data.  But, the same
picture leads to substantially different antishadowing for charged and
neutral current reactions, thus affecting the extraction of the
weak-mixing angle $\theta_W$ ~\cite{brodsky}.  This is due to the
fact that Reggeon couplings depend on the quantum numbers of the
struck quark implies non-universality of nuclear antishadowing for
charged and neutral currents ~\cite{brodsky}.

\section{New approaches to PDFs}

Moments of PDFs are matrix elements of local gauge invariant operators
which in principle can be calculated by using lattice QCD.  A brief
review of recent lattice effort in determining the PDFs was presented
at the workshop ~\cite{renner}.  Lattice QCD calculations of three
representative observables, the transverse quark distribution,
momentum fraction, and axial charge, were presented ~\cite{renner}. 
It was emphasized that lattice calculations of nucleon structure are
beginning to realize their promise to elucidate QCD and make contact
with the experimental programs.  It was concluded that recent
calculations are painting a qualitative three dimensional picture of
nucleon structure revealing a significant $x$ dependence of the
transverse size of the nucleon.  Quantitative calculations of moments
of PDFs are progressing, in particular, the calculation of $g_A$ may
soon reach a few percent accuracy. 

An analytical approach to understand the three dimensional picture of
nucleon structure was presented at the workshop ~\cite{belitsky}.  
A concept of the quantum phase-space (Wigner) distributions for the
quarks and gluons in the nucleon was introduced. The quark Wigner
functions were related to the transverse-momentum dependent PDFs
and generalized PDFs with emphasis on the physical role
of the skewness parameter.   Any knowledge on the generalized PDFs can
be immediately translated into the correlated coordinate and momentum
distributions of partons. In particular, the generalized PDFs can be
used to visualize the phase-space motion of the quarks, and hence
allow studying the contribution of the quark orbital angular momentum
to the spin of the nucleon.  It was concluded that 
measurements of generalized PDFs and/or direct lattice QCD
calculations of them will provide a fantastic window to the quark
and gluon dynamics in the proton ~\cite{belitsky}.

Another presentation made at the workshop looked at quark asymmetries
in nucleons ~\cite{alwall}.  Instead of fitting the data, a physical 
model for the non-perturbative $x$-shape of PDFs was developed.  The
model was based on Gaussian fluctuations in momenta, and quantum
fluctuations of the proton into meson-baryon pairs.  It was found
that the model gives a good description of the proton structure
function and a natural explanation of observed quark asymmetries, such
as the difference between the anti-up and anti-down sea quark
distributions and between the up and down valence distributions
~\cite{alwall}.  Within this model, there is an asymmetry in the
momentum distributions of strange and anti-strange quarks in the
nucleon, and the asymmetry is large enough to reduce the NuTeV anomaly
to a level which does not give a significant indication of physics
beyond the standard model.

Effective field theory was used to investigate the nuclear
modification to the PDFs and a recent result was presented at the
workshop ~\cite{detmold}. It was found that the universality of the
shape distortion in nPDFs (the factorization of the Bjorken $x$ and
atomic weight $A$ dependence) is model independent and emerges
naturally in effective field theory.  For a simple parameterization of
nonperturbative functions in the approach, fits to the data confirm
the factorization ~\cite{detmold}.


\begin{theacknowledgments}
We would like to thank all the members of our working group for the 
excellent presentations and for lively discussions they provoked.  We would
also like to thank the conveners of the Electroweak and Beyond the Standard
Model working group for their assistance in the joint session on high $Q^2$
structure function measurements.  We would also like to thank all the session
chairs for agreeing to be involved.  Last, but not least, we would like to
thank the organizers of DIS 2005 for interesting and well-organized meeting.
\end{theacknowledgments}


\bibliographystyle{aipproc}   

\end{document}